\documentclass[aps,twocolumn,showpacs]{revtex4}

\usepackage{graphicx}
\usepackage{dcolumn}
\usepackage{bm}

\begin{document}

\title{Delocalization in harmonic chains with
long-range correlated random masses}

\author{F.~A.~B.~F. de Moura, M.~D. Coutinho-Filho,
E.~P. Raposo}
\affiliation{Laborat\'orio de F\'{\i}sica Te\'orica e Computacional,
Departamento de F\'{\i}sica, Universidade Federal de Pernambuco,
50670-901 Recife, PE, Brazil}
\author{M.~L. Lyra}
\affiliation{Departamento de F\'{\i}sica, Universidade Federal de Alagoas,
57072-970 Macei\'o, AL, Brazil}

\begin{abstract}

We study the nature of collective excitations in harmonic chains with masses exhibiting
long-range correlated disorder with power spectrum
proportional to $1/k^{\alpha}$, where $k$ is the wave-vector
of the modulations on the random masses landscape.
Using a transfer matrix method and exact diagonalization, we compute
the localization length and  participation ratio of eigenmodes within the
band of allowed energies.  We find extended
vibrational modes in the low-energy region for $\alpha > 1$. In order to study
the time evolution of an initially localized energy input, we calculate  the
second moment $M_2(t)$ of the energy spatial distribution.
We show that $M_2(t)$, besides  being dependent of the specific
initial excitation and exhibiting an anomalous diffusion for
weakly
correlated disorder, assumes a ballistic spread in the regime $\alpha>1$
 due to the presence of extended vibrational modes.
\end{abstract}
\maketitle

\section{Introduction}

The role played by disorder on the nature
of collective excitations in condensed matter physics has been the
subject of intensive studies due to its relevance in defining general transport
characteristics~\cite{mack}. Usually, disorder induces localization of collective
excitations thus degrading transport properties, an effect that is largely pronounced
in low dimensions. In particular, the one-electron eigen-states in the
one-dimensional Anderson model with site-diagonal uncorrelated disorder are
exponentially localized for any degree of disorder~\cite{abrahams}.
However, several one-dimensional models with correlated
disorder have been proposed which exhibit
delocalized states~\cite{flores1,dunlap,evangelou1}.
Recently, it has been shown that the  one-dimensional Anderson model with
long-range correlated disorder presents a phase of extended
electronic states~\cite{chico,chico2,izrailev}. These results have been
confirmed by  microwave transmission spectra of  single-mode
waveguides with inserted correlated scatters~\cite{apl2}.

The above results have motivated the study of further model systems that can be mapped
onto the Anderson model and, therefore, expected to present a similar transition between
localized and extended collective excitations.
Recently, a study concerning the one-dimensional quantum Heisenberg
ferromagnet with exchange couplings exhibiting long-range correlated disorder  reported
some finite-size scaling evidences of the emergence of a phase of extended low-energy
excitations~\cite{delson}. By using a renormalization group calculation the existence of such phase of extended
spin-waves was confirmed and the scaling of the mobility edge with the degree
of correlation was obtained~\cite{mauricio}. It was also shown that, associated with the
emergence of extended spin-waves in the low-energy region, the wave-packet  mean-square
displacement exhibits a long-time ballistic behavior.

The collective vibrational motion of one-dimensional disordered
harmonic chains of $N$ random masses can also be mapped onto an
one-electron  tight-binding model~\cite{dean1}. In such a case, most of the normal
vibrational modes are localized.
However, there are a few low-frequency modes not localized,
whose number is of the order of $\sqrt{N}$,
in which case the disordered chains behaves
like the disorder-free system~\cite{dean1,matsuda}.
Futher, it was shown that correlations in the mass distribution produce a new set of non-scattered modes in this
system~\cite{datta1}. Non-scattered modes have also been found in disordered harmonic chain with
dimeric correlations in the spring constants~\cite{domi}.
By using analytical arguments, it was also
demonstrated that the transport of energy in mass-disordered (uncorrelated and correlated)
harmonic chains is strongly dependent on non-scattered vibrational modes as well as on
the initial excitation~\cite{datta2}. For
impulse initial excitations, uncorrelated random chains have a
superdiffusive behavior for the second moment of the energy distribution [$M_2(t) \propto
t^{1.5}$], while for initial displacement excitations a subdiffusive spread takes place
[$M_2(t) \propto t^{0.5}$]. The dependence of the second moment spread on the
initial excitation was also obtained in Ref.~\cite{wagner}.
Moreover, correlations induced by thermal annealing have been shown to enhance the localization
length of vibrational modes, although they still present an exponential decay for distances larger
than the thermal correlation length~\cite{lyra}. Recently the thermal conductivity on harmonic and anharmonic
chains of uncorrelated random masses~\cite{prl1}, as well as of the chain of hard-point particles of alternate masses~\cite{prl2},
has been numerically  investigated in detail. The main issue here  is whether the systems
display finite thermal conductivity in the thermodynamic
limit, a question that remains controversial~\cite{prl3}.

In  this paper we extend the study of collective modes in the
presence of long-range correlated disorder for the case of vibrational modes.
We will consider harmonic chains with long-range correlated random
masses assumed to have  spectral power density $S \propto 1/k^{\alpha}$.
By using a transfer matrix calculation, we obtain accurate estimates for the
Lyapunov exponent, defined as the inverse of the degree of localization $\lambda_c$ of
the vibrational modes. We show that, for $\alpha >1$, this model also presents a phase of extended
modes in the low frequency region. This result is confirmed by
participation
ratio measurements from an exact diagonalization procedure and finite size
scaling arguments. The spatial evolution of an initially localized excitation
is also studied by
computing the spread of the
second moment of the energy distribution, $M_2(t)$. We find that, associated with the
emergence of a phase of delocalized modes, a ballistic energy spread takes place.

\section{Formalism}

We consider a disordered harmonic chain of $N$ masses, for which the equation of motion
for the displacement $u_n$ of
the {\it n}-th mass with vibrational frequency $\omega$ is~\cite{matsuda,datta1}
\begin{equation}
(\beta_{n-1}+\beta_n-\omega^2
m_n)u_n=\beta_{n-1}u_{n-1}+\beta_nu_{n+1}~~~.
\end{equation}
Here $m_n$ is the mass at the {\it n}-th site and  $\beta_n$ is the spring constant that
couples the masses $m_n$ and $m_{n+1}$. We use units in which $\beta_n=1$. In the present
harmonic chain model, we take the masses $m_n$ following a random sequence describing the trace
of a fractional Brownian motion~\cite{feder,osborne,greis}:
\begin{equation}
m_n = \sum_{k=1}^{N/2}
 \left[k^{-\alpha }\left(\frac{2\pi }{N}\right)^{(1-\alpha )}\right]^{1/2}
\cos{\left( \frac{2\pi nk}{N} +\phi_k\right)},
\end{equation}
where $k$ is the wave-vector of the modulations on the random
mass landscape and $\phi_k$ are $N/2$ random phases uniformly distributed
in the interval $[0,2 \pi ]$. The exponent $\alpha$ is directly related to the
Hurst exponent $H$ ($\alpha =2H+1$) of the rescaled range analysis.
In order to avoid vanishing masses we shift and normalize
all masses generated by Eq.~(2) such to have average
value $<m_n>= 5$ and variance independent of the chain size
($<\Delta m_n> \equiv 1$).

Using the matrix formalism,  Eq.~(1) can be rewritten as
\begin{figure}
\centering
\includegraphics*[width=0.45\textwidth]{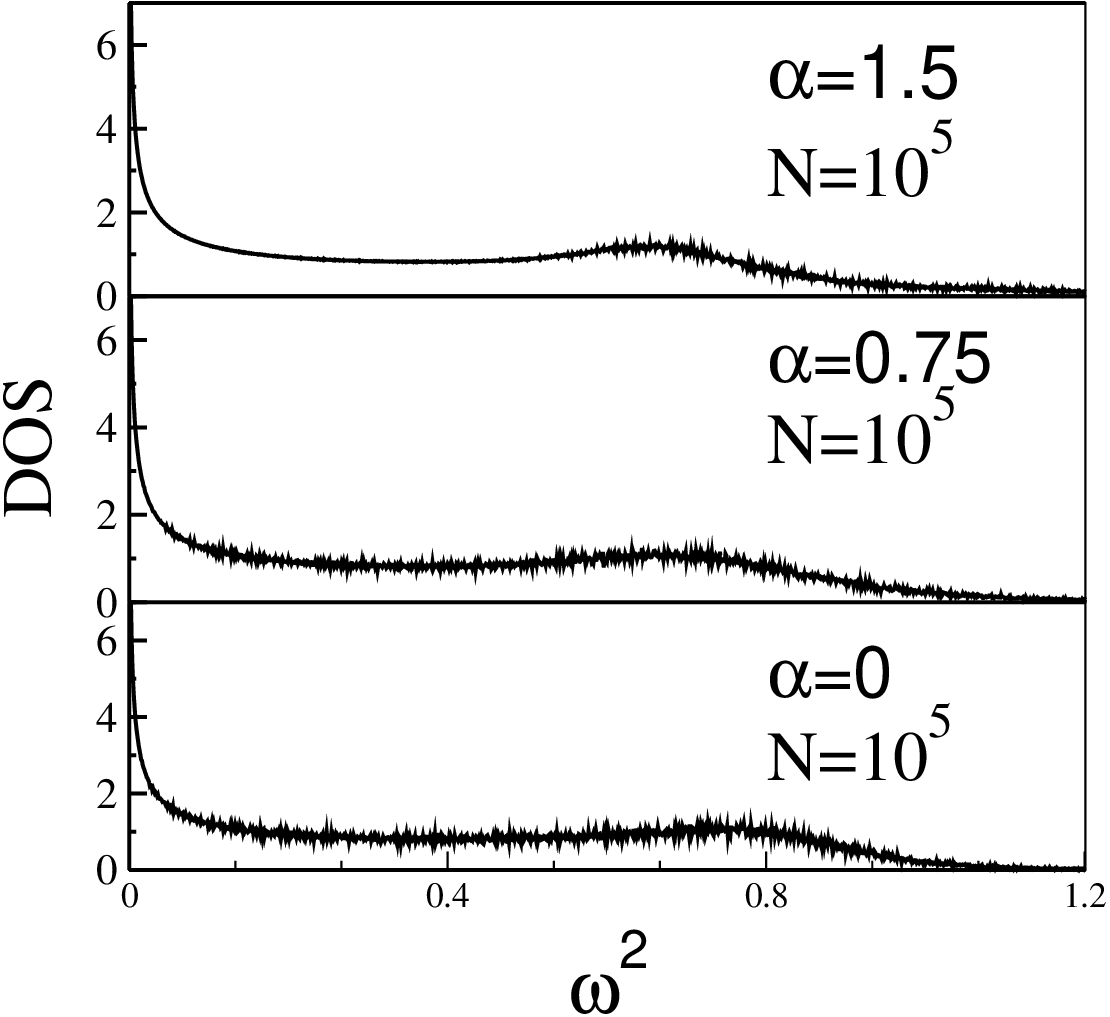}
\caption{Normalized density of states (DOS) as a function of $\omega ^2$
obtained using Dean's method. The chain length is $N=10^5$ for all cases.
The DOS becomes less rough as $\alpha$ is increased. For $\alpha=1.5$ it displays a
non-fluctuating part near the bottom of the band.}
\end{figure}
\begin{equation}
\left(\begin{array}{c}
u_{n+1} \\
u_n
\end{array}\right) =\left(\begin{array}{cc}
2-m_n\omega ^2 & -1 \\
1       &  0 \\
\end{array}\right) \left(\begin{array}{c}
u_n  \\
u_{n-1}
\end{array}\right) ~.
\end{equation}
For a specific frequency $\omega$, a $2 \times 2$ transfer matrix $T_n$ connects the
displacements at the sites $n-1$ and $n$ to those at the site $n+1$:
\begin{equation}
T_n=
\left(\begin{array}{cc}
  2-m_n\omega ^2 & -1 \\
  1     & 0
  \end{array}\right).
\end{equation}
Once the initial values for $u_0$ and $u_1$ are known, the value of $u_n$ can be
obtained  by
repeated iterations along the chain, as described by the product of transfer matrices
\begin{equation}
Q_N=\prod_{n=1}^{N}T_n.
\end{equation}
\begin{figure}
\centering
\includegraphics*[width=0.45\textwidth]{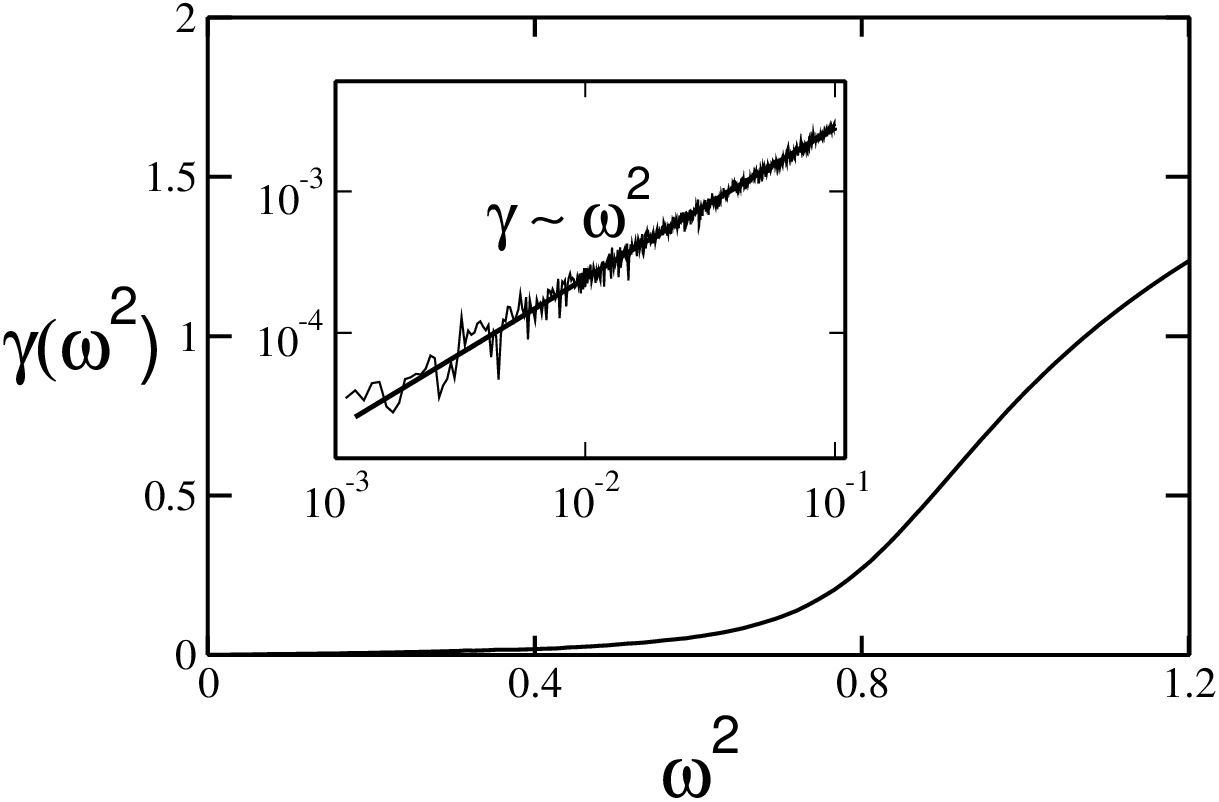}
\caption{Lyapunov coefficient $\gamma$ as a function of $\omega ^2$ for $\alpha = 0$
(uncorrelated  random chain) and $N=2 \times 10^5$ sites. The Lyapunov coefficient is finite for
non-zero frequencies (localized states)
and vanishes as $\gamma \propto \omega^2$, $\omega \rightarrow 0$ (inset).}
\end{figure}
The localization length of each vibrational mode is taken as
the inverse  of the Lyapunov exponent $\gamma$ defined by~\cite{matsuda,datta1,evangelou2}
\begin{equation}
\gamma = \lim_{N \to \infty}\frac{1}{N}\log \frac{|Q_Nc(0)|}{|c(0)|},
\end{equation}
where  $c(0)={u_1 \choose u_0}$ is a generic initial condition.
Typically, $2 \times 10^5$ matrix products were used to calculate the Lyapunov exponents.
The nature of the  vibrational modes can also be
investigated by computing the participation ratio $\xi$,
since it displays a dependence on the chain size for
extended states and is finite for exponentially localized ones.
$\xi$ is defined by~\cite{domi,lyra}
\begin{equation}
\xi(\omega) = \frac{\sum_{n=1}^N u_n^2}{\sum_{n=1}^N u_n^4},
\end{equation}
where the displacements $u_n$ are those associated with an
eigenmodes $\omega$ of a chain of $N$ masses and are obtained by direct diagonalization of
 the
$N \times N$ secular matrix $A$ defined by
$A_{i,i}=(\beta_i+\beta_{i-1})/m_i = 2/m_i$,
$A_{i,i+1}=A_{i+1,i}=\beta_i^2/(m_im_{i+1})=1/(m_im_{i+1})$, and
all other $A_{i,j}=0$~\cite{dean1,datta1}. The participation ratio calculations were averaged
over
$100$ samples.

We compute the Lyapunov exponents and the participation ratio
for several values of the correlation exponent $\alpha$,
and obtain the density
of states (DOS) using the numerical Dean's method~\cite{dean}. Strong fluctuations in
the DOS are  related to the presence of localized states,
whereas smooth a DOS is usually connected with the
emergence of delocalized states~\cite{evangelou1,datta1}.
In Fig.~1 we show the normalized DOS for chains with $N=10^5$ sites,
and notice that it becomes less rough as $\alpha$ is increased. In Fig.~2 we display the plot of
$\gamma$ versus $\omega^2$  for $\alpha =0$ (uncorrelated random masses). The Lyapunov coefficient
is finite for all frequencies and vanishes at $\omega=0$  as $\gamma \propto w^2$,
in agreement with Ref.~\cite{matsuda} (see inset of Fig.~2).  The scaled  participation ratio
$\xi/N$ as a function of $\omega ^2$ is shown in Fig.~3: $\xi (\omega^2=0)/N$  remains finite
in the thermodynamic limit, whereas for any finite frequency the vibrational modes are localized
with $\xi/N \rightarrow 0$ as $N \rightarrow \infty$, in agreement with the above results obtained
from the Lyapunov coefficient calculations. To investigate the effect of weak long-range correlated disorder,
we present in Fig.~4(a) the
Lyapunov coefficient as a function of $\omega ^2$ for $\alpha =0.75$ and  $N=2 \times 10^5$.
In spite of
$\gamma$ being very small in the bottom of the band, the scaled participation ratio for $\omega>0$
vanishes in the thermodynamic limit  [see Fig.~4(b)]. Therefore, all modes with $\omega>0$ are
still
localized, a feature that holds for any
$0 \leq \alpha \leq 1$. However, the nature of the low-frequency modes changes qualitatively for
$\alpha > 1$. In Fig.~5(a) we show $\gamma$ versus $\omega^2$  for $\alpha =1.5$
and $N=2 \times 10^5$ sites.
The Lyapunov coefficient vanishes within a finite range of frequency values, thus revealing the presence
of extended vibrational modes.  The  scaled
participation ratio $\xi/N$ [see  Fig.~5(b)] displays a well defined data collapse, confirming
that the phase of extended low-frequency
vibrational modes is stable in the thermodynamic limit.

\begin{figure}
\centering
\includegraphics*[width=0.45\textwidth]{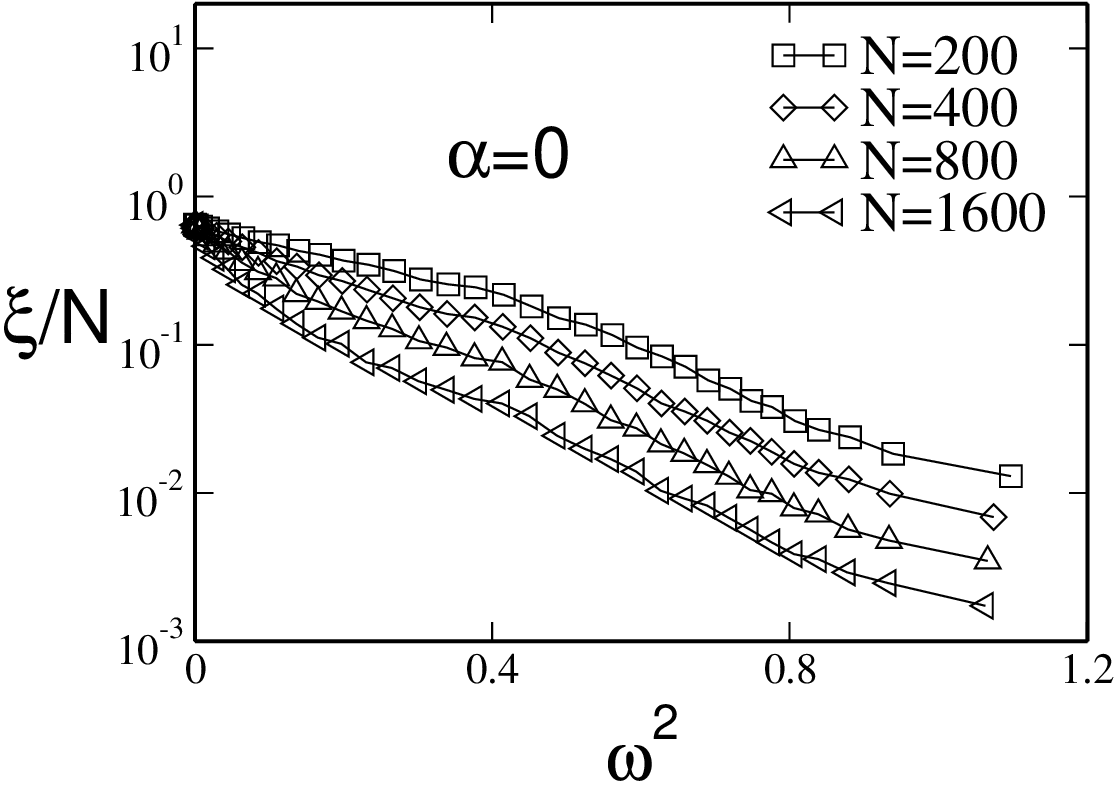}
\caption{Scaled  participation ratio $\xi/N$ as a function of $\omega ^2$
for $\alpha = 0$ (uncorrelated disorder).
From top to bottom, $N = 200, 400, 800, 1600$.
For vibrational modes with
$\omega > 0$, $\xi/N \rightarrow 0$ as $N$ diverges, thus confirming their localized nature.}
\end{figure}
\begin{figure}
\centering
\includegraphics*[width=0.45\textwidth]{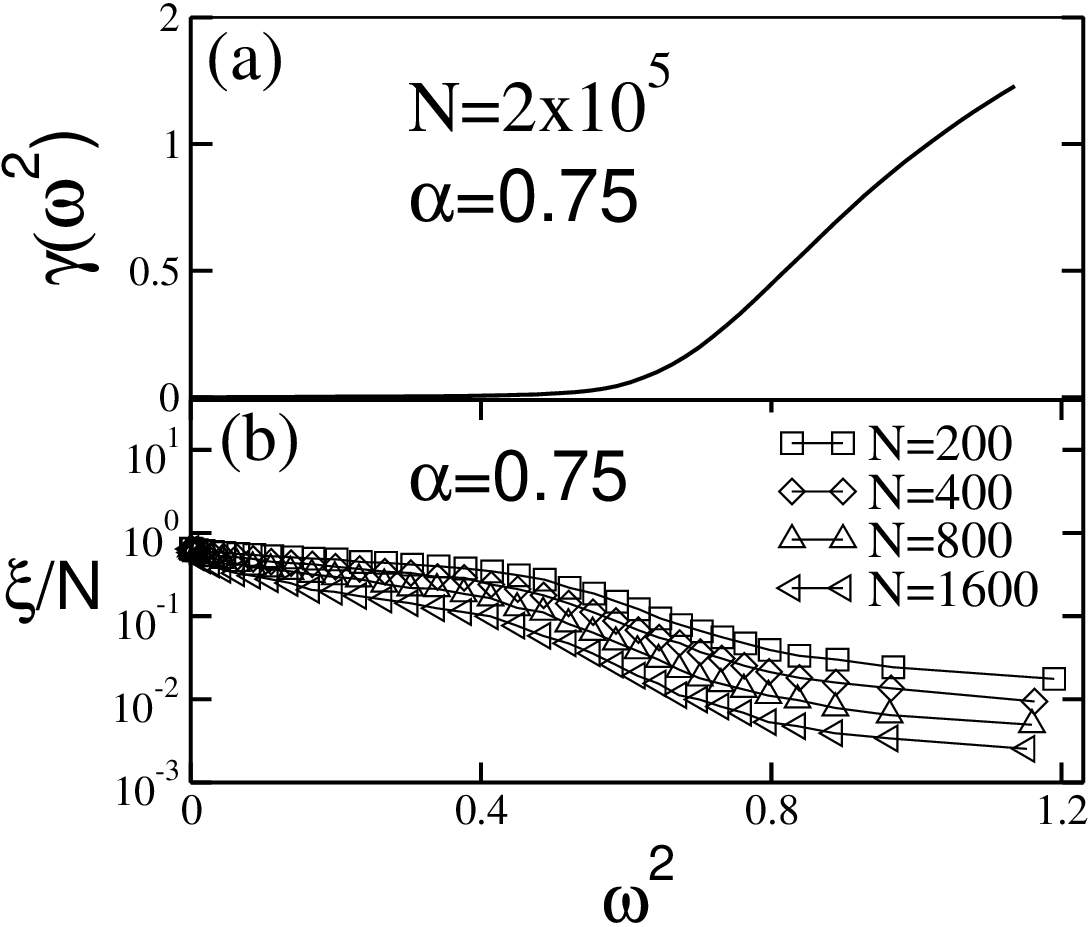}
\caption{(a)~Lyapunov coefficient $\gamma$ versus $\omega ^2$
for $\alpha = 0.75$ and $ N=2 \times 10^5$ sites. (b)~Scaled participation ratio $\xi/N$
as a function of $\omega ^2$ for $\alpha = 0.75$. From top to bottom,
$N=200, 400, 800, 1600$. In spite of
$\gamma$ being very small in the bottom of the band, all modes with $\omega>0$ are localized.}
\end{figure}
\begin{figure}
\centering
\includegraphics*[width=0.45\textwidth]{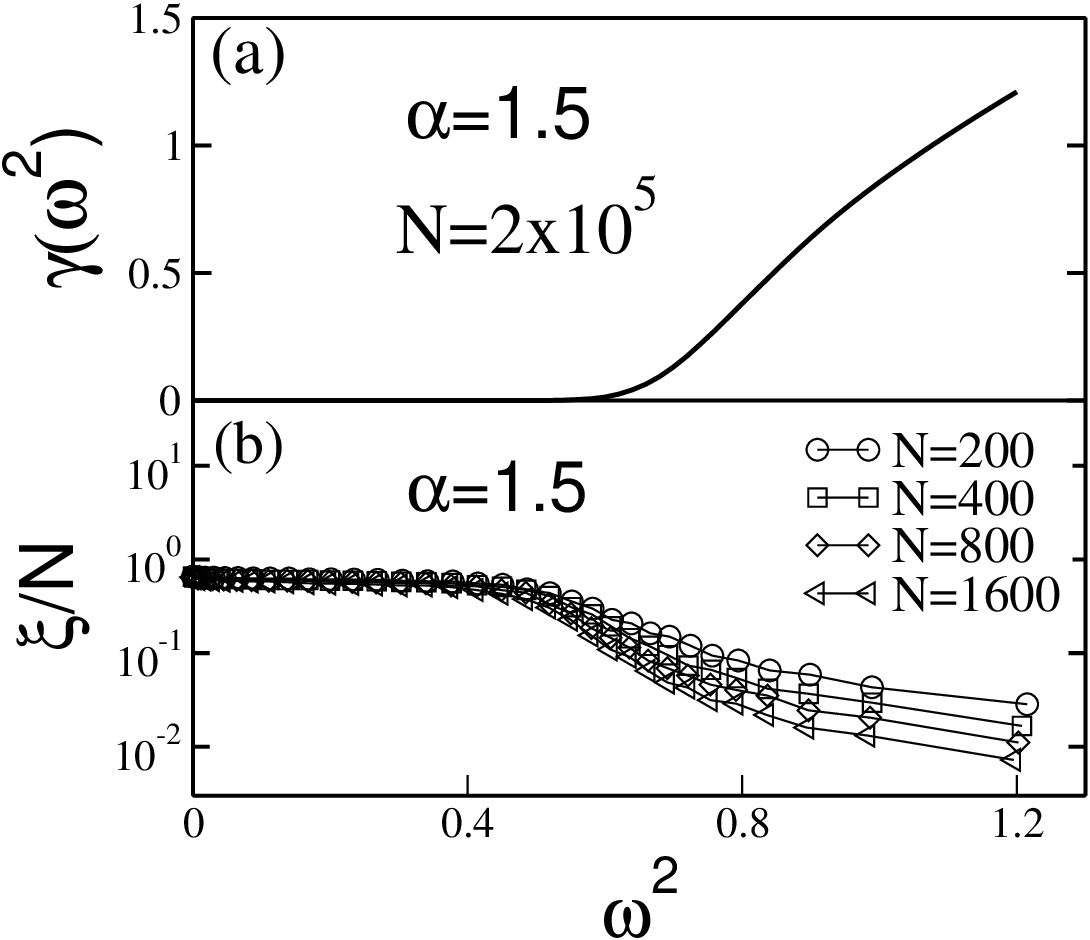}
\caption{(a)~Lyapunov coefficient $\gamma$ versus $\omega ^2$
 for $\alpha = 1.5$ and $ N=2 \times 10^5$ sites. The Lyapunov
coefficient vanishes within a finite range of frequency values, thus revealing the presence
of extended vibrational modes. (b)~Scaled  participation ratio $\xi/N$ as a function of $\omega ^2$ for $\alpha = 1.5$.
From top to bottom, $N=200, 400, 800, 1600$. The phase of extended vibrational modes is
confirmed by the
size independent plateau in the low-frequency region.}
\end{figure}

\section{Energy transport}

In order to study the time evolution of a localized energy pulse, we calculate
the second moment of the energy distribution~\cite{datta2,wagner}.  This quantity is related to the
thermal conductivity by Kubo's formula~\cite{kubo,datta2}. The classical Hamiltonian $H$ for an harmonic chain
can be written as
\begin{equation}
H=\sum_{n=1}^N h_n(t)~,
\end{equation}
where the energy $h_n(t)$ at the site $n$ is given by
\begin{equation}
h_n(t)=\frac{P_n^2}{2m_n}+\frac{\beta_n}{4}[(Q_{n+1}-Q_n)^2+(Q_{n}-Q_{n-1})^2]~.
\end{equation}
Here $P_n$ and $Q_n$ define the momentum and displacement of the mass at the {\it n}-th site.
The Hamilton's equations are
\begin{equation}
\dot{P}_n(t)= -\frac{\partial H}{\partial Q_n}=\beta_n [(Q_{n+1}-Q_n)-(Q_{n}-Q_{n-1})]
\end{equation}
and
\begin{equation}
\dot{Q}_n(t)=\frac{\partial H}{\partial P_n}=\frac{P_n(t)}{m_n}~.
\end{equation}

The fraction  of the total energy $H$ at the
site $n$ is given by $h_n(t)/H$ and the second moment of
the energy distribution, $M_2(t)$, is defined by~\cite{datta2}
\begin{equation}
M_2(t)=\sum_{n=1}^N (n-n_0)^2 [h_n(t)/H],
\end{equation}
\begin{figure}
\centering
\includegraphics*[width=0.45\textwidth]{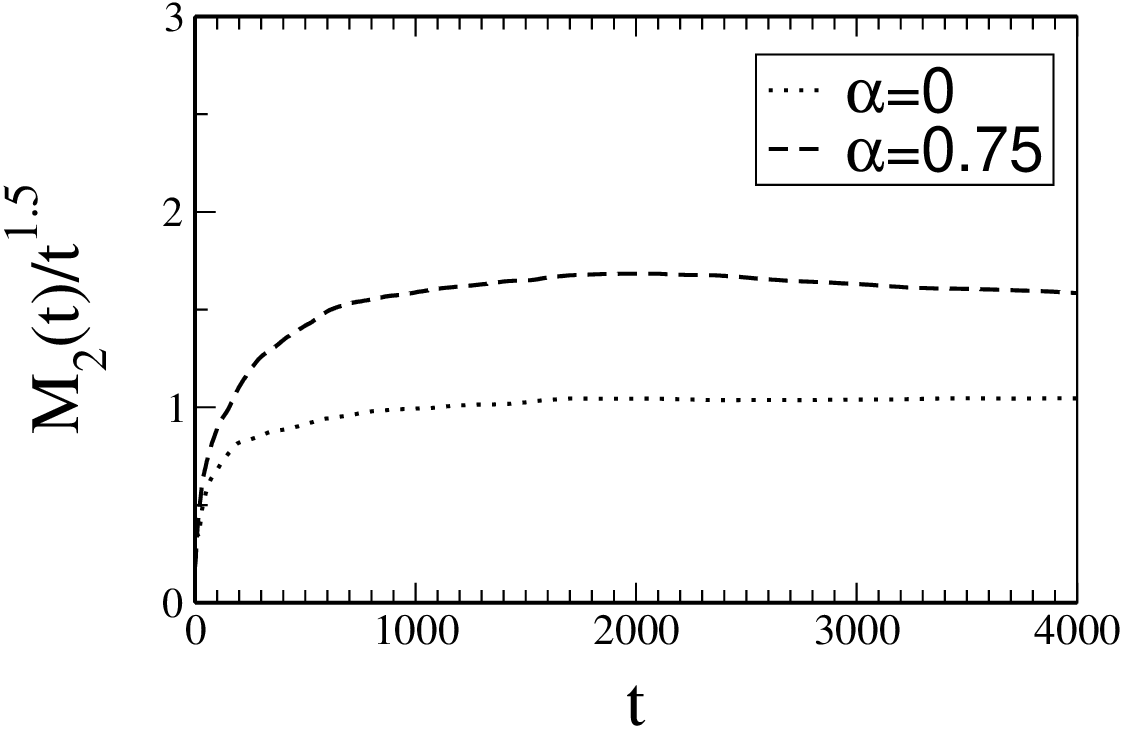}
\caption{Scaled energy second moment $M_2(t)/t^{1.5}$ versus time $t$ for
$\alpha=0$ (dotted line) and
$\alpha=0.75$ (dashed line) with initial impulse excitation. For
$0 \leq \alpha \leq 1$ only superdiffusive behavior
is found for long times.}
\end{figure}

\noindent
where an initial excitation is introduced at the site $n_0$ at $t=0$ . Using the fourth-order
Runge-Kutta method, we solve the differential equations for $P_n(t)$ and $Q_n(t)$
and calculate $M_2(t)$.  The second moment of the energy distribution $M_2(t)$ has the same status
of the mean-square displacement of the wavepacket of an electron in a crystal~\cite{datta2}.
In harmonic chains with an initial impulse excitation, the energy spread
is faster than that in chains with an initial displacement excitation~\cite{datta2,wagner}.
We calculate $M_2(t)$ for several $\alpha$ values and two kinds
of initial excitation: impulse excitation and displacement excitation.
\subsection{Impulse excitation}
In Fig.~6 we present the scaled second moment $M_2(t)/t^{1.5}$
versus time $t$ for $\alpha=0$ (dotted line), which corresponds to the uncorrelated random chain, and
$\alpha=0.75$ (dashed line).
These results have been obtained after
an initial impulse excitation, $P_{n_0}(t=0)=\delta_{n_0,N/2}$.
In our
calculations for $\alpha=0$,
the self-expanded chain method with initial chain size $N=1000$ was used  to minimize end effects.
Throghout the numerical integration process
we kept the fraction of the total energy $H$ at the ends of the chain [$h_0(t)/H$ and  $h_N(t)/H$]
smaller than $10^{-300}$ for all times.
As shown in Fig.~6, we find a long-time superdiffusive behavior for $\alpha = 0$,
in agreement with previous analytical
and numerical results
for energy transport in harmonic chains with uncorrelated random masses under an
impulse initial excitation~\cite{datta2}.
In contrast, for  $\alpha > 0$ we cannot
use the self-expanded chain method due to the long-range character of the mass correlations.
Therefore, chains with $N=10000$ masses were considered, and the runs stopped whenever the
fraction of the total energy at the chain ends achieved $10^{-300}$.
For $\alpha = 0.75$ the time-dependence of the scaled energy second moment $M_2(t)/t^{1.5}$
typically represents a weak long-range correlated case.
In such a case we also find superdiffusive behavior for long times.
On the other hand,
in the strong correlated regime, $\alpha > 1$, a breakdown in the superdiffusive behavior sets up.
Fig.~7 shows the time-dependence of the scaled energy second moment, $M_2(t)/t^{2}$, for
$\alpha=1.5$ (dotted line) and $\alpha=2.0$ (dashed line). Associated with the
emergence of extended vibrational modes in the low-energy region,
the second moment $M_2(t)$ displays a long-time ballistic behavior.
\begin{figure}
\centering
\includegraphics*[width=0.45\textwidth]{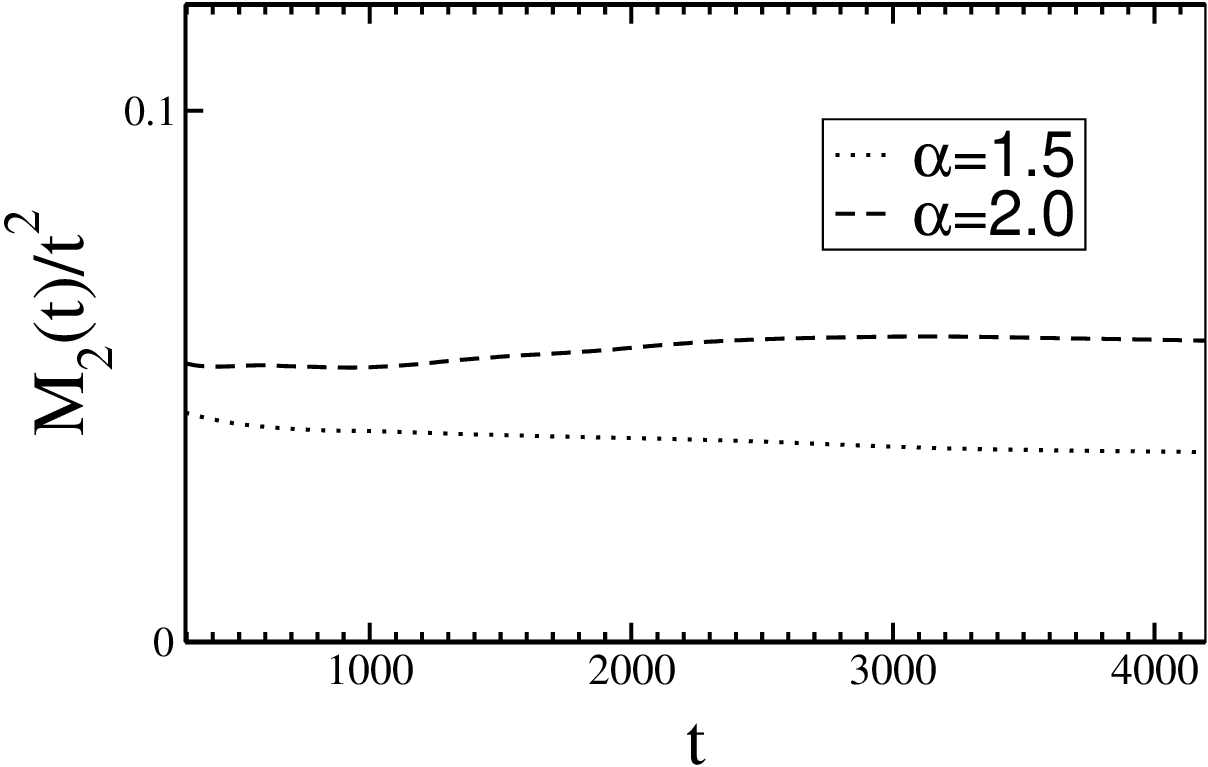}
\caption{Scaled second moment $M_2(t)/t^2$ versus time $t$ for $\alpha=1.5$ (dotted line) and
$\alpha=2.0$ (dashed line) with initial impulse excitation. Results were obtained by numerical integration
in chains with $N=10000$ sites. A ballistic behavior
is found after an initial transient.}
\end{figure}
\subsection{Displacement excitation}
The long time behavior of the second moment $M_2(t)$ in uncorrelated random chains
with initial displacement excitation is significantly different from the
corresponding behavior with impulse initial excitation~\cite{datta2,wagner}.
Analytical calculations
predict that  $M_2(t)\propto t^{0.5}$, a result
that has been corroborated by numerical techniques~\cite{datta2}.
For $\alpha=0$  we
indeed
reproduce this behavior, as shown in  Fig.~8 for the scaled second moment $M_2(t)/t^{0.5}$ versus time $t$ with initial
displacement excitation $Q_{n_0}(t=0)=\delta_{n_0,N/2}$. Again, we find that
this
 asymptotic subdiffusive behavior remains true
 for $0 \leq \alpha <1$ (dashed line in Fig.~8).
For strong correlations ($\alpha > 1$), which  induce the emergence of
new extended vibrational modes in the low-energy region, the
energy transport is faster than in the subdiffusive regime and again assumes a ballistic nature,
as shown in Fig.~9.
\begin{figure}
\centering
\includegraphics*[width=0.45\textwidth]{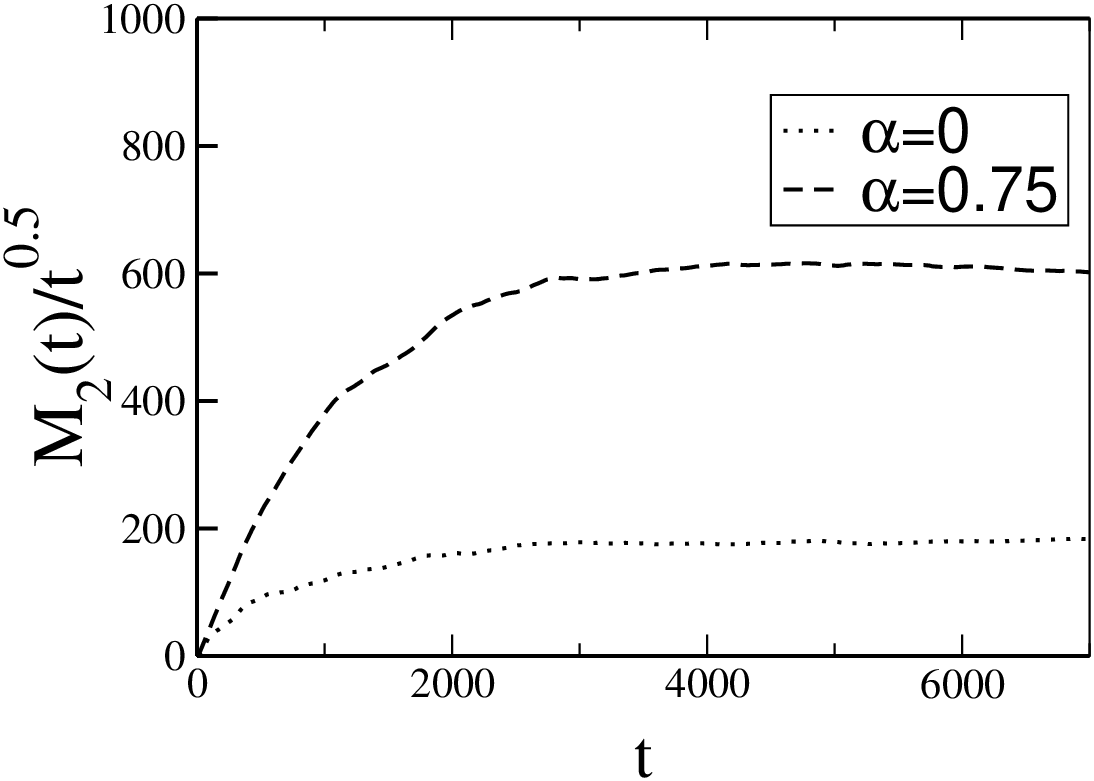}
\caption{The scaled second moment $M_2(t)/t^{0.5}$ versus time $t$ for $\alpha=0$(dotted line) and
$\alpha=0.75$ (dashed line) with initially given displacement excitation.  The subdiffusive behavior
is found for long times.}
\end{figure}

\section{Summary and Conclusions}

In  this paper we have studied the nature of collective
vibrational modes in harmonic chains with long-range
correlated random masses $m_n$, with spectral power
density $S \propto 1/k^{\alpha}$. By using a transfer matrix method
and exact diagonalization, we have computed the localization length and
the participation ratio of all normal modes. Our results indicate  that
in the strong correlations regime,  $\alpha>1$,  there is a phase
of extended low-energy vibrational modes. In this sense, long-range correlations
in the mass distribution induce the emergence of a delocalization
transition in harmonic chains similar to the one observed to occur with
one-magnon excitations in ferromagnetic chains with random couplings~\cite{delson,mauricio}
and with one-electron
eigen-states in the random hopping Anderson model~\cite{chico2}.
We have also studied the energy transport
in this harmonic chain  model. The  spread  of the energy
second moment $M_2(t)$ is shown to be strongly dependent on
the existence of non-scattered vibrational modes and initial excitation. We have
also found that, associated with the emergence of a phase of low-energy
extended collective excitations, $M_2(t)$ displays a crossover from
an anomalous sub- or super-diffusive regime (depending on the
initial impulse or displacement excitation, respectively)
to an asymptotic ballistic behavior. The above findings indicate that
the thermal conductivity can be strongly influenced by the presence of long-range
correlations in the random distribution of masses  and we hope that the present
work will stimulate further studies along this direction.

\section{Acknowledgments}

This work was partially supported by CNPq, CAPES and FINEP (Brazilian agencies).
MLL also acknowledges the partial support of FAPEAL (Alagoas state agency).

\begin{figure}
\centering
\includegraphics*[width=0.45\textwidth]{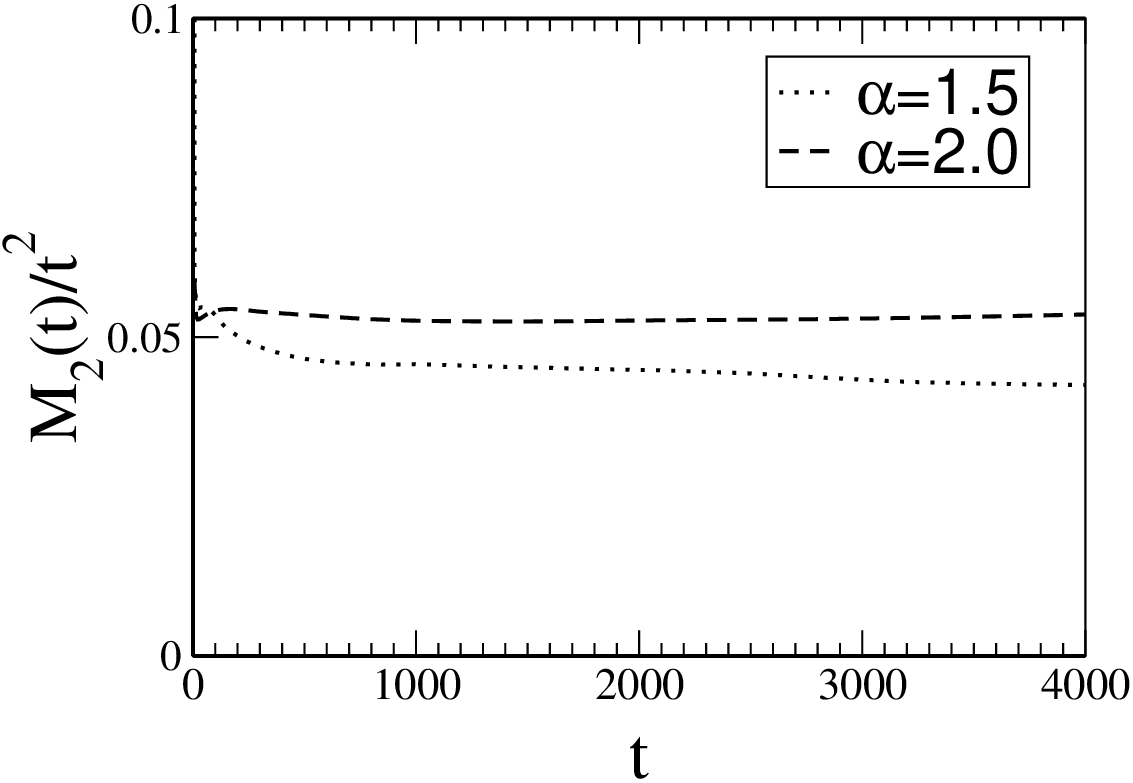}
\caption{The scaled second moment $M_2(t)/t^{2}$ versus time $t$ for $\alpha=1.5$ (dotted line) and
$\alpha=2.0$ (dashed line) with
initially given displacement excitation. Results obtained by numerical integration of diferential equation for chain
with $N=10000$ sites. A ballistic behavior is found for all times.}
\end{figure}

\newpage

\noindent

\end{document}